\documentclass[12pt]{article}
\usepackage{amsmath}
\usepackage{amssymb,amsthm}
\usepackage{color}

\newtheorem{Thm}{Theorem}[section]
\newtheorem{theorem}[Thm]{Theorem}
\newtheorem{proposition}[Thm]{Proposition}
\newtheorem{corollary}[Thm]{Corollary}
\newtheorem{lemma}[Thm]{Lemma}

\newtheorem{remark}{Remark}[section]

\newtheorem{definition}[Thm]{Definition}

\title{A theory of quantum (statistical) measurement}

\author{Walter F. Wreszinski\footnote{wreszins@gmail.com, 
Instituto de Fisica, Universidade de S\~ao Paulo (USP), Brasil}}        
        
\begin{document}

\maketitle

\begin{abstract}
We propose a theory of quantum (statistical) measurement which is close, in spirit, to Hepp's theory, which is centered 
on the concepts of decoherence and macroscopic (classical) observables, and apply it to a model of the Stern-Gerlach experiment. 
The number $N$ of degrees of freedom of the measuring apparatus is such that $N \to \infty$, justifying the adjective
``statistical'', but, in addition, and in contrast to Hepp's approach, we make a three-fold assumption: the measurement is not 
instantaneous, it lasts a finite amount of time and  is, up to arbitrary accuracy, performed in a finite region of space, 
in agreement with the additional axioms proposed by Basdevant and Dalibard. It is then shown how von Neumann's ``collapse postulate'' 
may be avoided by a mathematically precise formulation of an argument of Gottfried, and, at the same time, Heisenberg's 
``destruction of knowledge'' paradox is eliminated. The fact that no irreversibility is attached to the process of measurement is shown 
to follow from the author's theory of irreversibility, formulated in terms of the mean entropy, due to the latter's property of affinity.
\end{abstract}

\section{Introduction and Summary}

In a recent very stimulating paper, S. Doplicher \cite{Dop} describes qualitatively a ``possible picture of the measurement process in
quantum mechanics, which takes into account the finite and nonzero time duration $T$ of the interaction between the 
observed system and the microscopic part of the measurement apparatus''. In this paper we do not distinguish, as he does, two
parts of the measurement apparatus, which, for us, will be a ``macroscopic pointer'', modelled by a quantum system with number
of degrees of freedom $N= \infty$, as suggested by Hepp \cite{Hepp}, but the time-duration $T$ of the measurement will be assumed
to satisfy the conditions:
\begin{itemize}
\item [$a.)$] $0 < T$; 
\item [$b.)$] the measurement takes place in a region of finite spatial extension
\end{itemize} 

In their quantum mechanics textbook for the \'{E}cole Polytechnique, Basdevant and Dalibard \cite{BasDal} remark, in connection with
their analysis of the Stern-Gerlach (SG) experiment \cite{StGer}, that a.) and b.) are ``two fundamental aspects which are absent
from the classical formulation of the principles of quantum mechanics''.

In his concluding remarks in section 3, Doplicher observes that ``the conventional picture of the measurement process in quantum mechanics''
requires that, as $N \to \infty$, the time duration of the measurement tends to zero and that the measurement apparatus occupies a volume
$V$ such that $V \to \infty$, referring in this context to the important work of Araki and Yanase \cite{AY}. The latter authors also show,
however, for a simple case, that an approximate measurement of an operator such as spin is possible to any desired accuracy. A similar
result follows, in our approach, which relies in the framework introduced by Haag and Kastler \cite{HK}, by restriction to a class of 
observables which are ``arbitrarily close to their restriction to finite $N$''(corresponding to finite volume, assuming finite density, 
as required in the thermodynamic limit) - see Assumption A in section 2. In this sense, b.) above will follow, as in the case examined
by Araki and Yanase, to arbitrary accuracy. 

Concerning, however, the requirement that the time duration tend to zero, the situation is completely different, at least in a 
nonrelativistic context (in the relativistic field context, the same should follow for entirely different reasons, see the conclusion).
Our forthcoming Theorem 3.4 strongly requires assumption a.), i.e., that the measurement not be instantaneous,
and, in the concrete SG model of section 4, it may be explicitly seen that if $T(N) \to 0$ at a certain rate (see (82) of Remark 4.1),
the off-diagonal elements of the density matrix do not vanish as $N \to \infty$. We explain why we are not forced to require that the
time of measurement be instantaneous in Remark 4.2: it has to do with the forthcoming notion of macroscopic or classical observables.
In section 5 we shall also see that preparation of the system
and measurement are dual, inseparable processes, and in the hypothesis of their both being instantaneous, a ``time-arrow'' may not
exist a priori, which is an essential condition for a precise formulation of the author's condition of irreversibility \cite{Wre}.

Doplicher's choice of conventional picture of the measurement process, the article \cite{DLP}, in his view ``quite satisfactory'', has, in
our opinion (as well as Hepp's, see (\cite{Hepp}, p. 243)), one major disadvantage: it employs, in a crucial sense, the ``ergodic average'',
which is not supported by any physical principle.

We now briefly describe our framework, following, in part, \cite{Dop}. In von Neumann's general picture \cite{vN}, we have a system $S$,
whose general observable $A=\sum_{j} \lambda_{j} E_{j}$ has finite spectrum $\lambda_{j}, j=1, \cdots, n$, and self-adjoint spectral projections
$E_{j}$. The Hilbert space of the state vectors of the composite system, consisting of $S$ and the measurement apparatus $A_{N}$, which we assume
to consist of a quantum system with $N$ degrees of freedom, is given by the tensor product ${\cal H}_{S} \otimes {\cal H}_{A_{N}}$ of the
corresponding Hilbert spaces. The total Hamiltonian is
\begin{equation}
\label{(1.3)}
H_{N} = H_{S} \otimes \mathbf{1} + \mathbf{1} \otimes H_{A_{N}} + V_{N}
\end{equation}
For simplicity, we restrict further the number of eigenvalues of the observable $A$ to two, $\lambda_{+}$ and $\lambda_{-}$, with 
$\lambda_{+} > \lambda_{-}$ (as will be the case in the SG experiment of section 4). There exists a quantity $t_{D}$, called 
\emph{decoherence time} (or relaxation time), which may be explicitly computed in the SG model, defined as the minimum time interval $t_{D}$
such as a measurement of $A_{N}$, i.e., such that $\lambda_{+}$ and $\lambda_{-}$ may be experimentally distinguished, is possible. We assume
that 
\begin{equation}
\label{(1.4)}
0 < t_{D} \mbox{ and } t_{D} \mbox{ is independent of } N
\end{equation}
Our requirement on $T$, compatible with assumption a.), may be stated as
\begin{equation}
\label{(1.5)}
0 < t_{D} \le T < \infty \mbox{ with } t_{D} \mbox{ and } T \mbox{ independent of } N
\end{equation}

In an important paper, Narnhofer and Thirring \cite{NTh1} examined the intriguing question why the only states found in Nature are such
that they assume definite values on classical observables, but never mixtures of them. This problem has been lively discussed since 
Schr\"{o}dinger introduced his cat \cite{Schr}. As simple examples of classical (or macroscopic) observables, they propose the mean
magnetization of a magnet
\begin{equation}
\label{(1.6)}
\vec{m} = \lim_{N \to \infty} \frac{1}{2N+1} \sum_{i=-N}^{N} \vec{\sigma_{i}}
\end{equation}
or the center of mass velocity of a system of particles
\begin{equation}
\label{(1.7)}
\vec{v} = \lim_{N \to \infty} \frac{\sum_{i=-N}^{N} m_{i} \vec{v}_{i}}{\sum_{i=-N}^{N} m_{i}}
\end{equation}
of a large object. We shall use both in this paper, but replace \eqref{(1.7)} by the center of mass coordinate of a particle system
\begin{equation}
\label{(1.8)}
\vec{x}_{C.M.} = \lim_{N \to \infty} \frac{1}{2N+1} \sum_{i=-N}^{N} \vec{x}_{i}
\end{equation}
(of a group of equal atoms). 
 
We remark that  \eqref{(1.6)} - \eqref{(1.8)} are precise definitions of macroscopic or classical observables when one specifies the
appropriate representation, as we do in section 2. It is in this connection that the limit $N \to \infty$ plays a crucial role in the
present framework and, in this respect, quite analogously to Hepp's \cite{Hepp}.

In order to explain the problems, we adopt Bell's suggestion (\cite{Bell1}, p.36) of taking the apparatus $A_{N}$ out of the ``rest of
the world'' R, and treat it together with $S$ as part of the enlarged quantum system $S_{N}^{'}$: $R=A_{N}+R^{'}$;$S+A_{N}=S_{N}^{'}$;
$W=S_{N}^{'}+R^{'}$: ``the original axioms about 'measurement' are then applied not at the $S/A_{N}$ interface, but at the $A_{N}/R^{'}$
interface''. Neglecting the interaction of $A_{N}$ with $R^{'}$, the joint system $S_{N}^{'}$ is found to end, by the Schr\"{o}dinger
equation associated to $H_{N}$ in \eqref{(1.3)}, after the ``measurement on $S$ by $A_{N}$'' (i.e., after a fixed time $T$ satisfying 
\eqref{(1.5)})in a state
\begin{equation}
\label{(1.9)}
\Psi_{N}(T) = \sum_{n}c_{n} \Psi_{n,N}(T)
\end{equation}
where the states $\Psi_{\pm,N}(T)$ correspond to two definite (apparatus) pointer positions. The corresponding density matrix is
\begin{equation}
\label{(1.10.1)}
\rho_{N}(T) = \sum_{n,m} c_{n}\bar{c_{m}} \Psi_{n,N}(T)\overline{\Psi_{m,N}(T)}
\end{equation}
where the bar denotes complex conjugation.
Bell reports that in his textbook analysis of the measurement problem, Kurt Gottfried (\cite{KG}, pp. 186-188) insists that, being
$A_{N}$ a macroscopic system (and thus also $S_{N}^{'}$), 
\begin{equation}
\label{(1.10.2)}
tr(A\hat{\rho})=tr(A\rho) \mbox{ ``for all observables $A$ known to occur in Nature'' }
\end{equation}
where 
\begin{equation}
\label{(1.11)}
\hat{\rho}_{N}(T) = \sum_{n} |c_{n}|^{2} \Psi_{n,N}(T) \overline{\Psi_{n,N}(T)}
\end{equation}
(in our notation) - ``dropping interference terms involving pairs of macroscopically different states''. We shall refer to the replacement
of $\rho_{N}(T)$ by $\hat{\rho}_{N}(T)$ as the ``von Neumann collapse of the density matrix''. The associated ``loss of relative phases''
leads to what we shall refer to as \emph{Heisenberg paradox} \cite{Heis}: ``Every experiment destroys some of the knowledge of the system
which was obtained by previous experiments''. We shall see that, while a reduction
of type \eqref{(1.10.2)}, \eqref{(1.11)} does not occur for finite $N$, it may indeed occur in the limit $N \to \infty$: this is the
content of Corollary 3.5, which makes the last sentence in \eqref{(1.10.2)} precise, i.e., specifies the (physically sensible) class of
observables $A$. This enables elimination of one of Bell's objections in \cite{Bell2} to Hepp's conceptual framework: the observable
which ``undoes the measurement'' proposed by him does not exist in the specified framework, see \cite{NarWre}. On the contrary, his
second objection in \cite{Bell2}, that the infinite-time limit in the only example of automorphic evolution considered by
Hepp, the Coleman model, is not physically sensible, is sound. Indeed, this model does \emph{not} satisfy \eqref{(1.5)}, because
$$
t_{D} = t_{D}(N) = N+ \mbox{ constant }
$$
where $N$ denotes the number of sites in the model's (spin) chain (\cite{Se1}, \cite{Se2}, \cite{NarWre}): thus $t_{D}(N) \to \infty$
as $N \to \infty$. It thus turns out that Bell's criticism applies to the model, rather than to the whole conceptual framework introduced
by Hepp and, indeed, Narnhofer and Thirring provide a physically reasonable model example in which the infinite time limit can be 
controlled and agrees with some of Hepp's conclusions (\cite{NTh1}, see their Remark 1). This example is, however, not very illuminating
from the point of view of measurement theory, having being designed to describe certain interactions with the environment which render
a mixed state pure in the infinite time limit, while we are interested in the opposite effect, that a pure state becomes mixed under
evolution. For this reason, we analyse in section 4 a model of the SG experiment, which well illustrates Theorem 3.4 and is a generalization
to an infinite number of degrees of freedom of the model proposed in \cite{KG1}, together with the prescription of initial state and 
experimental setting in \cite{GG}, see also \cite{BasDal}.

The states in the assumption of theorem 3.4 depend on the parameter $T$, which is only supposed to satisfy \eqref{(1.5)}.
Concerning this point, the idea should be mentioned (\cite{Bell1}, p.37, bottom) that 
``systems such as $S_{N}^{'}$ have \emph{intrinsic} properties - independently of and before observation''. For instance, the ``jump'' 
associated to the collapse is supposed to occur at some not well specified time (\cite{Bell2}, p. 98).
However, both the Landau-Lifshitz-Bohr-Haag picture of measurement as an interaction between system $S_{N}^{'}$ and environment
$R^{'}$ which occurs apart from and independently of any observer (\cite{LL}, \cite{Haag}), as well as the fact, emphasized by
Peierls \cite{Pe1} that the observer does not have to be contemporaneous with the event, allowing, for example, from present
evidence, to draw conclusions about the early Universe (the classical example being the cosmic microwave background), strongly 
suggest that the quantities to be measured do not depend on $T$. Ideally, we expect that the states in Theorem 3.4 satisfy 
the assumptions of the theorem \emph{for all} $T$ satisfying \eqref{(1.5)}, and, moreover, that the actually measured quantities 
independ of $T$. It is rewarding that the example treated in section 4 fulfills both of these expectations (see Remark 4.2).

In section 5 we briefly review the definition of irreversibility in (\cite{Wre}, \cite{Wre1}) in terms of the mean entropy \cite{LanRo}, and 
prove that it is conserved on the average under ``collapse'', as a consequence of the property of affinity \cite{LanRo}. This result 
contrasts with Lemma 3 of \cite{NarWre}, where the quantum Boltzmann entropy of a finite system is shown to decrease under collapse, thus
contradicting the second law (on the average), and requiring that the incidence of interactions with the environment be rare 
in order to assure the global validity of the second law (see the last remarks in \cite{NarWre}). As a consequence of theorem 5.1,
van Kampen's conjecture (\cite{vK}, mentioned in \cite{Bell1}) that the entropy of the Universe remains zero throughout the process 
of measurement is confirmed \emph{in the sense of the mean entropy}, and thus the ``irreversibility paradox'' suggested by 
Landau and Lifschitz \cite{LL} and Gottfried \cite{KG} does not take place for infinite quantum spin systems, adopting the
mean entropy as indicator. An illustration of Theorem 5.1 in the theory of measurement is provided by the effective quantum spin model 
of the SG experiment in section 4.2.

Section 6 is reserved to a conclusion, with a brief discussion of open problems.

The present paper owes very much to the theory of quantum statistical mechanics of infinite systems, as described in \cite{BRo2}, with
a pedagogical textbook exposition in the classic book by Sewell \cite{Se}. The basic Theorem 3.4 amalgamates results in the papers of
Roberts and Roepstorff \cite{RRoe} and Hepp \cite{Hepp}. The groundbreaking framework of the paper of Haag and Kastler \cite{HK},
nicely reviewed by Wightman \cite{Wight} plays a central role in the proposed framework. 

Concerning references, a good bibliography on several aspects of the quantum theory of measurement up to 2003 is to be found in 
\cite{KG1}, pp. 575 and 576. Several other recent references, including book references, may be found in \cite{Dop}. From the point
of view of mathematical physics, a very recent reference is \cite{Fro}: there, it is argued that the Schr\"{o}dinger equation does
not yield a correct description of the quantum mechanical time evolution of states of isolated physical systems featuring events; it
also cites several recent references, to which we refer. In a different framework, that of thermal open systems, a recent reference is
\cite{Pil}, see also references given there.

In the introduction and elsewhere, we sometimes state ``we assume...'': in order to clarify what is really assumed, we have collected
\emph{all} the assumptions in Assumption A in section 2. 

\section{General setting}

\subsection{Generalities: states of infinite systems}

We very briefly summarize here some concepts of crucial importance in this paper, but, for any detail, we refer to the references (\cite{Se}, 
\cite{BRo2}, \cite{Hug}). we shall use quantum spin systems as a prototype, such as the generalized Heisenberg Hamiltonian
\begin{equation}
\label{(2.1)}
H_{\Lambda} = -2\sum_{x,y \in \Lambda}[J_{1}(x-y)(S_{x}^{1}S_{y}^{1}+S_{x}^{2}S_{y}^{2})+J_{2}(x-y)S_{x}^{3}S_{y}^{3}]
\end{equation}
where
\begin{equation}
\label{(2.2)}
\sum_{x \in \mathbf{Z}^{\nu}}|J_{i}(x)|< \infty \mbox{ and } J_{i}(0)=0 \mbox{ for } i=1,2
\end{equation}
Above, $\vec{S}_{x} \equiv (S_{x}^{1},S_{x}^{2},S_{x}^{3})$, where $S_{x}^{i}=1/2 \sigma_{x}^{i}, i=1,2,3$ and $\sigma_{x}^{i}, i=1,2,3$ are the
Pauli matrices at the site $x$. Above, $H_{\Lambda}$ acts on the Hilbert space ${\cal H}_{\Lambda}=\otimes_{x \in \Lambda}\mathbf{C}_{x}^{2}$,
and $\vec{S}_{x}$ is short for $\mathbf{1} \otimes \cdots \otimes \vec{S}_{x} \otimes \cdots \otimes \mathbf{1}$. The algebra associated
to a finite region $\Lambda \subset \mathbf{Z}^{\nu}$ is
\begin{equation}
\label{(2.3)}
{\cal A}(\Lambda) = B({\cal H}_{\Lambda})
\end{equation}
and two of its properties are crucial:
\begin{itemize}
\item [$a.)$] (causality)$[{\cal A}(B),{\cal A}(C)]=0$ if $B \cap C = \phi$; 
\item [$b.)$] (isotony) $B \subset C \Rightarrow {\cal A}(B) \subset {\cal A}(C)$.
\end{itemize}
\begin{equation}
\label{(2.6)}
{\cal A}_{L} = \cup_{B} {\cal A}(B)
\end{equation}
where $B$ ranges over the finite parts of $\mathbf{Z}^{\nu}$,
is called the \emph{local} algebra; its closure with respect to the norm 
\begin{equation}
\label{(2.7)}
{\cal A} \equiv \overline{{\cal A}_{L}}
\end{equation}
is the \emph{quasilocal} algebra: it consists of observables which are, to arbitrary accuracy, approximated by observables 
attached to a \emph{finite} region. The bar in \eqref{(2.7)} denotes the C*-inductive limit (\cite{KR}, Prop.11.4.1).
The norm is defined by $A \in B({\cal H}_{\Lambda}) \to ||A|| = sup_{||\Psi|| \le 1} ||A \Psi||$, $\Psi \in {\cal H}_{\Lambda}$.
An \emph{automorphism} one-to one mapping of ${\cal A}$ into ${\cal A}$ which preserves the algebraic structure:
$A \to \tau_{x}(A)$ denotes the space-translation automorphism.

A \emph{state} $\omega_{\Lambda}$ on ${\cal A}(\Lambda)$ is a positive, normalized linear functional on ${\cal A}(\Lambda)$:
$\omega_{\Lambda}(A) = Tr_{{\cal H}_{\Lambda}} (\rho_{\Lambda} A) \mbox{ for } A \in {\cal A}(\Lambda)$
(positive means $ \omega_{\Lambda}(A^{\dag}A) \ge 0$, normalized $\omega_{\Lambda}(\mathbf{1})=1$.)

For quantum spin systems, the index $N$ will be identified as
\begin{equation}
\label{(2.19d)}
N = |\Lambda| = V
\end{equation}
with the understanding that $N \nearrow \infty$ means, for simplicity, the limit along a sequence of parallelepipeds of sides
$a_{i}, i=1, \cdots, \nu$, with $a_{i} \to \infty$ for each $i \in [1,\nu]$; more general limits, such as the van Hove limit
(\cite{BRo2}, p. 287) could be adopted.

The notion of state generalizes to systems with
infinite number of degrees of freedom $\omega(A)= \lim_{\Lambda \nearrow \infty} \omega_{\Lambda}(A)$, at first for $A \in {\cal A}_{L}$
and then to ${\cal A}$.

Each state 
$\omega$ defines a representation $\Pi_{\omega}$ of ${\cal A}$ as bounded operators on a Hilbert space ${\cal H}_{\omega}$ with
cyclic vector $\Omega_{\omega}$ (i.e., $\Pi_{\omega}({\cal A}) \Omega_{\omega}$ is dense in ${\cal H}_{\omega}$), such that 
$\omega(A) = (\Omega_{\omega}, \Pi_{\omega}(A) \Omega_{\omega})$ (the GNS construction). The strong closure of $\Pi_{\omega}({\cal A})$ is
a von Neumann algebra, with commutant $\Pi_{\omega}({\cal A})^{'}$, which is the set of bounded operators on ${\cal H}_{\omega}$ which
commute with all $\Pi_{\omega}({\cal A})$, and the center is defined by $Z_{\omega}= \Pi_{\omega}({\cal A}) \cap \Pi_{\omega}({\cal A})^{'}$.

The set of atates over the algebra ${\cal A}$ will be denoted by $E_{{\cal A}}$.

Considering quantum spin systems on $\mathbf{Z}^{\nu}$, we shall consider only space-translation-invariant states, i.e., such that
\begin{equation}
\label{(2.8)}
\omega \circ \tau_{x} = \omega \mbox{ for all } x \in \mathbf{Z}^{\nu}
\end{equation} 

An extremal invariant or ergodic state is a state which cannot be
written as a proper convex combination of two distinct states $\omega_{1}$ and $\omega_{2}$, i.e., the following does \emph{not}
hold:
\begin{equation}
\label{(2.9)}
\omega = \alpha \omega_{1} + (1-\alpha) \omega_{2} \mbox{ with } 0<\alpha<1
\end{equation}
If the above formula is true, it is natural to regard $\omega$ as a mixture of two pure ``phases''
$\omega_{1}$ and $\omega_{2}$, with proportions $\alpha$ and $1-\alpha$, respectively (\cite{BRo1}, Theorem 2.3.15).

A \emph{factor} or \emph{primary} state is defined by the condition that the center
\begin{equation}
\label{(2.10)}
Z_{\omega}= \{\lambda \mathbf{1} \}
\end{equation}
with $\lambda \in \mathbf{C}$.

For quantum spin systems the center $Z_{\omega_{\beta}}$ coincides (\cite{BRo1}, Example 4.2.11) with the so called algebra at infinity
$\zeta_{\omega}^{\perp}$, which corresponds to operations which can be made outside any bounded set. As a typical example of an observable
in $\zeta_{\omega}^{\perp}$, let $\omega$ be any translation invariant state. Then the space average of A
\begin{equation}
\label{(2.11a)}
\eta_{\omega}(A) \equiv s-lim_{\Lambda \nearrow \infty} \frac{1}{|\Lambda|} \sum_{x \in \Lambda} \Pi_{\omega}(\tau_{x}(A))
\end{equation}
exists, and, if $\omega$ is ergodic, then 
\begin{equation}
\label{(2.11b)}
\eta_{\omega}(A) = \omega(A) \mathbf{1}
\end{equation}
(\cite{LvH}), which corresponds to ``freezing''
the observables at infinity to their expectation values. The following definition is abstracted from \cite{Hepp}, before his Lemma 1.

\begin{definition}
\label{Definition 2.1}
Two states $\omega_{1}$ and $\omega_{2}$ are \emph{disjoint} if no subrepresentation of
of $\Pi_{\omega_{1}}$ is unitarily equivalent to any subrepresentation of $\Pi_{\omega_{2}}$. Two states which induce disjoint 
representations are said to be disjoint; if they are not disjoint, they are called \emph{coherent}.
\end{definition}

For finite-dimensional matrix algebras (with trivial center) all representations are coherent, and factor representations as well.

We have (\cite{Hepp}, Lemma 6): Let $\omega_{1}$ and $\omega_{2}$ be extremal invariant (ergodic) states with respect to space
translations. If, for some $A \in {\cal A}$,
\begin{equation}
\label{(2.11c)}
\eta_{\omega_{1}} (A) = a_{1} \mbox{ and } \eta_{\omega_{2}}(A) = a_{2} \mbox{ with } a_{1} \ne a_{2}
\end{equation}
then $\omega_{1}$ and $\omega_{2}$ are disjoint.

The space averages $\eta$ defined above correspond to macroscopic ``pointer positions'', e.g., the mean magnetization in the
Heisenberg model \eqref{(2.1)} in the $3$- direction $\sum_{x \in \Lambda} \frac{S_{x}^{3}}{|\Lambda|}$, with $A= S^{3}$. 
If $\eta_{\omega_{+}}(S^{3}) = a_{+} = 1$, and $\eta_{\omega_{-}}(S^{3}) = -1$, the states $\omega_{\pm}$ are macroscopically
different, i.e., differ from one another by flipping an infinite number of spins. For a comprehensive discussion, see 
\cite{Se}, section 2.3.

Given a state $\omega_{1}$, the set of states $\omega_{2}$ ``not disjoint from'' $\omega_{1}$ forms a \emph{folium}: a norm-closed
subset ${\cal F}$ of $E_{{\cal A}}$ such that
(i) if $\omega_{1},\omega_{2} \in {\cal F}$, and $\lambda_{1}, \lambda_{2} \in \mathbf{R}_{+}$ with $\lambda_{1}+\lambda_{2}=1$, then
$\lambda_{1} \omega_{1}+\lambda_{2} \omega_{2} \in {\cal F}$; ii.) if $\omega \in {\cal F}$ and $A \in {\cal A}$, the state $\omega_{A}$,
defined by 
\begin{equation}
\label{(2.12)}
\omega_{A}(B) = \frac{\omega(A^{*}BA)}{\omega(A^{*}A)} \mbox{ with } \omega(A^{*}A) \ne 0
\end{equation}
also belongs to ${\cal F}$ and is interpreted as a ``local perturbation of $\omega$''.

We shall denote the folium associated to a state $\omega$ by $[\omega]$. If two states $\omega_{1}$ and $\omega_{2}$ are disjoint, their
folia $[\omega_{1}]$ and $[\omega_{2}]$ are also disjoint. This follows from Hepp's Lemma 1 \cite{Hepp}:

\begin{lemma}
\label{lem:1}

$\omega_{1} \in E_{{\cal A}}$ and $\omega_{2} \in E_{{\cal A}}$ are disjoint if and only if for every representation $\pi$ of ${\cal A}$
with $\omega_{i} = \omega(\Psi_{i}) \circ \pi$ for some $\Psi_{i} \in {\cal H}_{\pi}$, $i=1,2$, one has
$$
(\Psi_{1}, \pi(A) \Psi_{2}) = 0 \forall A \in {\cal A}
$$

\end{lemma}

Above, $\omega_{i} = \omega(\Psi_{i}) \circ \pi$ means
$$
\omega_{i}(A) = (\Psi_{i}, \pi(A) \Psi_{i}) \mbox{ with } \Psi_{i} \in {\cal H}_{\pi}
$$
where ${\cal H}_{\pi}$ is the Hilbert space associated to the representation $\pi$. The lemma is easy to understand from the definition
~\ref{Definition 2.1} of disjointness: $\Psi_{2}$ and $\Psi_{1}$ lie in non-unitarily equivalent (``orthogonal'') Hilbert spaces, which
generally differ by different values of a macroscopic observable of type, e.g., (4), (5) or (6), which means an operation affecting an
\emph{infinite} number of points or sites, and therefore cannot be connected by a quasilocal observable, which is, by definition, 
arbitrarily close (in norm) to one localized in a finite region. The lemma also shows explicitly that when two states $\omega_{1}$ and
$\omega_{2}$ are disjoint, so are their folia, by definition ~\eqref{(2.12)}.
 
One important example, which will be our main concern in sections 4 and 5, is that of an infinite direct product space. For each vector
$\vec{m}_{i}$, with $\vec{m}_{i}^{2}=1$, there exists a vector $|\vec{m}_{i})$ in the Hilbert space $\mathbf{C}^{2}_{i}$ such that
$(\vec{\sigma}_{i} \cdot \vec{m}_{i})|\vec{m})_{i} = |\vec{m})_{i}$. Let ${\cal A}$ act on a reference vector \cite{NTh1}
$|\Psi_{\vec{m}}) = \otimes_{i=-\infty}^{\infty} |\vec{m})_{i} \mbox{ with } \vec{\sigma}_{i} |\vec{m}_{i}) = \vec{m} |\vec{m})_{i}$
For $\vec{m} \ne \vec{n}$, this yields two representations $\pi_{\vec{m}}, \pi_{\vec{n}}$ of ${\cal A}$ on separable Hilbert spaces 
${\cal H}_{\vec{m}}, {\cal H}_{\vec{n}}$. The following weak limits exist in these representations:
\begin{equation}
\label{(2.13a)}
\vec{m} \mathbf{1} = wlim_{N \to \infty} \frac{1}{2N+1} \sum_{i=-N}^{N} \pi_{\vec{m}}(\vec{\sigma}_{i})
\end{equation}
\begin{equation}
\label{(2.13b)}
\vec{n} \mathbf{1} = wlim_{N \to \infty} \frac{1}{2N+1} \sum_{i=-N}^{N} \pi_{\vec{n}}(\vec{\sigma}_{i})
\end{equation}
These two representations cannot be unitarily equivalent because
\begin{equation}
\label{(2.14)}
U^{-1} \pi_{\vec{m}}(\vec{\sigma}_{i}) U = \pi_{\vec{n}}(\vec{\sigma}_{i})
\end{equation}
would imply $U^{-1} \vec{m}\mathbf{1} U = \vec{n}\mathbf{1}$, which is impossible because $U$ cannot change the unity $\mathbf{1}$. 
The same argument shows disjointness. The
$\Psi_{\\vec{m}}$ define states $\omega_{\vec{m}}(\cdot) = (\Psi_{\vec{m}}, \cdot \Psi_{\vec{m}})$. The \emph{mixed} state is defined as
\eqref{(2.9)} (with $\vec{m} \ne \vec{n}$)
\begin{equation}
\label{(2.15)}
\omega_{\alpha} \equiv \alpha \omega_{\vec{m}} + (1-\alpha) \omega_{\vec{n}} \mbox{ with } 0 \le \alpha \le 1
\end{equation}
which is a convex combination of distinct pure states $\omega_{\vec{m}}$ and $\omega_{\vec{n}}$.
 
Consider, now, the framework described in section 1, consisting of the system $S$, for simplicity a spin one-half system, 
whose general observable is 
\begin{equation}
\label{(2.16)}
A = \lambda_{+}P_{+}+\lambda_{-}P_{-}
\end{equation}
Consideration of a general, finite spectrum of $A$ poses, however, no problem.
The Hilbert space of state vectors of the composite system will consist of $S$ and the measurement apparatus $A_{N}$, and
is given by the tensor product 
\begin{equation}
\label{(2.17)}
{\cal H}_{S} \otimes {\cal H}_{A_{N}}
\end{equation}
of the corresponding Hilbert spaces. The total Hamiltonian is
\begin{equation}
\label{(2.18)}
H_{N} = H_{S} \otimes \mathbf{1} + \mathbf{1} \otimes H_{A_{N}} + V_{N}
\end{equation}
We assume that later the limit $N \to \infty$ is taken in an appropriate sense. Take as initial state vector
\begin{equation}
\label{(2.19a)}
\Psi_{N}(t=0) = (\alpha |+) + \beta |-)) \otimes \Psi_{0}^{N}
\end{equation}
We assume that
\begin{equation}
\label{(2.19b)}
\exp(-iTH_{N}) \Psi_{N}(t=0) = \alpha |+) \otimes \Psi^{N,+,T} + \beta |-) \otimes \Psi^{N,-,T}
\end{equation}
with
\begin{equation}
\label{(2.19c)}
|\alpha|^{2} + |\beta|^{2} = 1
\end{equation}

\subsection{The framework: some specific assumptions}

We shall assume that the case of particle systems \eqref{(1.8)} is also included, replacing $\mathbf{Z}^{\nu}$ by $\mathbf{Z}$
and finite regions $\lambda$ by $\Lambda_{N} = [-N,N], N \in \mathbf{N}_{+}$, with $|\Lambda|=|\Lambda_{N}|=2N+1$ (see \eqref{(2.19d)}).
The isotony property b.) enables the algebra ${\cal A}$ associated to the apparatus to be defined as inductive limit
\eqref{(2.6)} (for the infinite product case, see \cite{Tak}). The algebra of the (system + apparatus) is thus assumed to be the
C*-inductive limit of the ${\cal A}_{s} \otimes {\cal A}_{\Lambda}$, denoted by 
\begin{equation}
\label{(2.20)}
{\cal A}_{s} \otimes {\cal A} 
\end{equation}
where ${\cal A}_{s}$ is the spin algebra, generated by the Pauli operators $\{\vec{\sigma}, \mathbf{1}\}$. Under assumption 
\eqref{(2.19b)}, we may, for each $T$ satisfying \eqref{(1.5)}, consider the states on ${\cal A}$
\begin{equation}
\label{(2.21)}
\omega_{\Lambda}^{+,T} = (\Psi^{N,+,T}, A \Psi^{N,+,T})
\end{equation}
and
\begin{equation}
\label{(2.22)}
\omega_{\Lambda}^{-,T} = (\Psi^{N,-,T}, A \Psi^{N,-,T})
\end{equation}
where
$$
A \in {\cal A}_{\Lambda}
$$

It is now convenient to distinguish explicitly the two cases we shall consider:

1.) Quantum spin systems

The natural topology (from the point of view of physical applications) in the space of states is the \emph{weak* topology}.
A sequence of states $\omega_{n}, n=1,2, \cdots$ on a C* algebra ${\cal A}$ is said to tend to a state $\omega$ in the weak* topology if
\begin{equation}
\label{(2.23)}
\lim_{n \to \infty} \omega_{n}(A) = \omega(A) \mbox{ for all } A \in {\cal A}          
\end{equation}

The above definition requires that we extend $\omega_{\Lambda}^{\pm,T}$ to ${\cal A}$ in one of the various possible ways,
for instance, assigning to the extension $\tilde{\omega}_{\Lambda}^{\pm,T}$ the value $1$ in the complement 
${\cal A}- {\cal A}_{\Lambda}$. Considering ${\cal A}$ as a
Banach space, since the set of states on ${\cal A}$ is sequentially compact in the weak*-topology
(see \cite{Roy}, Prop. 13, p.141 and Cor. 14, p. 142), because ${\cal A}$ is separable, there exists
a subsequence $\{\Lambda_{n_{k}}\}_{k=1}^{\infty}$ of $\Lambda_{n} \nearrow \infty$ and states $\tilde{\omega}^{\pm,T}$ on
${\cal A}$ such that 
\begin{equation}
\label{(2.24)}
\tilde{\omega_{k}}^{\pm,T}(A) \equiv \tilde{\omega}_{\Lambda_{n_{k}}}^{\pm,T}(A) \to \tilde{\omega}^{\pm,T}(A) \mbox{ as } k \to \infty
\end{equation}

2.) Particle systems.

In this case, we confine our attention to infinite product states on the infinite tensor product of C* algebras
$\otimes_{i \in \mathbf{Z}} {\cal A}_{i}$. Good references are \cite{Wehrl}, \cite{Gui}. In the sequel, take the index set 
$I = \mathbf{Z}$, and each ${\cal A}_{i}$,
with $i \in I$ to be the von Neumann algebra generated by the Weyl operators (for simplicity in one dimension, which
will be the case in the application in section 4)
$$
W(\beta, \gamma) = \exp[i(\beta z_{i}+ \gamma p_{z_{i}})]
$$
where $p_{z}=-i\frac{d}{dz}$, on ${\cal H}_{i}$ a copy of $L^{2}(\mathbf{R})$, with $\beta$ and $\gamma$ real numbers. 

\begin{definition}
\label{Definition 2.2}
Let $({\cal H}_{i})_{i \in I}$ be a family of Hilbert spaces. A family of vectors $(x_{i})_{i \in I}$, with $x_{i} \in {\cal H}_{i}$
is called a $C$ family if $\prod_{i \in I} ||x_{i}||$ converges. $(x_{i})_{i \in I}$ is called a $C_{0}$ family if
$\sum_{i \in I} |||x_{i}||-1|$ converges.
\end{definition}
It may be proved (see, e.g., \cite{Wehrl}, lemma 2.2) that every $C_{0}$ family is a $C$ family, and that every $C$ family 
fulfilling $\prod_{i \in I} ||x_{i}|| \ne 0$ is a $C_{0}$ family.

\begin{definition}
\label{Definition 2.3}
Two $C_{0}$ families $(x_{i})_{i \in I}$, $(y_{i})_{i \in I}$ are \emph{equivalent},$(x_{i})_{i \in I} \equiv (y_{i})_{i \in I}$, if
\begin{equation}
\label{(2.25)}
\sum_{i \in I} |(x_{i}|y_{i}) - 1| < \infty
\end{equation}
\end{definition}

It may be proved (see, e.g., \cite{Wehrl}, p. 60) that $\equiv$ is indeed an equivalence relation. The complete tensor product (CTP)
of the ${\cal H}_{i}$, denoted by $\otimes_{i \in I} {\cal H}_{i}$, defined in \cite{Wehrl}, p. 65, is a direct sum of 
\emph{incomplete tensor product spaces} (IDPS) $\otimes_{i \in I}^{\zeta} {\cal H}_{i}$: they are the closed linear subspaces of
the CTP spanned by the nonzero $C_{0}$ vectors in the $C_{0}$ family $\zeta$. If $0 \ne \otimes_{i \in I} x_{i} \in \zeta$, we write
$\otimes_{i \in I}^{(\otimes x_{i})_{i \in I}} {\cal H}_{i}$ for the IDPS. The important result for us in this connection will be

\begin{proposition}
\label{prop:2.1}
Let $\otimes_{i \in I} x_{i}$ be a $C_{0}$ vector not equal to zero. The set of all $\otimes_{i \in I} y_{i}$ such that $x_{i}=y_{i}$
for all but at most finitely many indices is total in $\otimes_{i \in I}^{\otimes_{i \in I} x_{i}} {\cal H}_{i}$.
\end{proposition}

(For a proof, see \cite{Wehrl}, p. 67, Prop. II.4).

In the application in section 4 we shall have states on an infinite tensor product of C* algebras ${\cal A}_{i}, i \in I$ (see
\cite{Gui}, p. 17, 2.2), which may also be defined as an inductive limit (\cite{Gui}, p. 18; \cite{Tak}) and will be denoted
by ${\cal A}$. For each $i \in I$, let $\omega_{i}$ be a state on ${\cal A}_{i}$, $\pi_{i}$ the associated GNS representation
(\cite{BRo1}, 2.3.3), with cyclic vector $\xi_{i}$.

\begin{definition}
\label{Definition 2.4}
The (infinite) product state $\otimes_{i \in I} \omega_{i}$ is the unique state on ${\cal A}$ verifying
\begin{equation}
\label{(2.26)}
(\otimes \omega_{i}) (\otimes x_{i}) = \prod \omega_{i}(x_{i}) \mbox{ for } x_{i} \in {\cal A}_{i}
\end{equation}
and $x_{i} = e_{i}$ for almost all $i$, where $e_{i}$ is the identity on ${\cal A}_{i}$.
\end{definition}

The representation of ${\cal A}$ canonically
associated to $\otimes_{i \in I} \omega_{i}$ is equivalent to the representation $\pi = \otimes_{i \in I} ^{\otimes_{i} \xi_{i}} \pi_{i}$
of ${\cal A}$ on $\otimes_{i \in I}^{\otimes \xi_{i}} {\cal H}_{i}$ such that $\pi(\otimes x_{i}) = \otimes \pi_{i}(x_{i})$ for
$x_{i} \in {\cal A}_{i}$, and $x_{i} = e_{i}$ for almost all $i$, where $e_{i}$ is the identity on ${\cal A}_{i}$.

(See Proposition 2.5, p. 20 and Proposition 2.9, p. 23, of \cite{Gui}).

We are now in the position of formulating our assumption - Assumption A - which will be the hypothesis of our main theorem (Theorem 3.4):

\emph{Assumption A}

Assume the framework consisting of the system S, for simplicity a spin one-half system with general observable \eqref{(2.16)}, and
Hamiltonian and initial state vector given by \eqref{(2.19b)}, under condition \eqref{(2.19c)}. 

In this connection, we also assume condition \eqref{(1.5)}.

The states $\tilde{\omega}^{\pm,T}$ of
quantum spin systems are defined by \eqref{(2.24)} with the algebra ${\cal A}$ (of the apparatus alone, appearing in \eqref{(2.20)}. 
For particle systems the initial state vector \eqref{(2.19a)} and those $\Psi^{M,\pm,T}$ at time $T$ in \eqref{(2.19b)} are vectors 
$\otimes_{i=-M}^{M} \xi_{i}^{\pm,T}$ with $M$ finite, and corresponding states $\omega_{M}^{\pm,T}$, while the states of the infinite 
system are the infinite product factor states $\tilde{\omega}^{\pm,T} \equiv \otimes_{i \in \mathbf{Z}} \omega_{i}^{\pm,T}$ of 
~\ref{Definition 2.4}, with corresponding factorial representation
$\otimes_{i \in \mathbf{Z}}^{\xi_{i}^{\pm,T}}\pi_{i}$. The algebra is ${\cal A}$, with ${\cal A}$ the infinite tensor product of
C* algebras. In each case, for all $A \in {\cal A}$ and given $\epsilon > 0$, there exists a finite positive integer $k$ and a strictly local
$A(\Lambda_{k}) = \pi_{k}(A)$, or an element $A_{k} = \pi_{k}(A)$ of $\otimes_{i=-k}^{k} {\cal A}_{i}$ such that
\begin{equation}
\label{(2.27a)}
||A - A(\Lambda_{k})|| < \epsilon
\end{equation}
or
\begin{equation}
\label{(2.27b)}
||A - A_{k}|| < \epsilon
\end{equation} 

\begin{remark}
\label{Remark 2.1}

In Assumption A, $\pi_{k}(A)$, for $A \in {\cal A}$ denotes a representation of ${\cal A}$ on a Hilbert space ${\cal H}_{\Lambda_{k}}$ (or ${\cal H}_{k}$
associated to the restriction of $A$ either to a local region or to a system with a finite number of particles, viz. satisfying \eqref{(2.27a)}. This
follows by construction, using the inductive limit structure of ${\cal A}$.

\end{remark} 

As a last remark, Assumption A is not so special as it might look: the way states of infinite systems are naturally obtained is precisely as
limits of finite systems, which actually describe the physical situation(s), in the natural weak* topology \eqref{(2.23)}. 

\section{General framework and main theorem}

Roberts and Roepstorff \cite{RRoe} have described a natural general framework for quantum mechanics, which includes systems with an
infinite number of degrees of freedom. Since their building blocks are, just as in the previous subsection, the algebra of observables
${\cal A}$ and the states $\omega$, we are able to adapt it to the present context in a very simple way, which we now describe.

We assume that $k=1,2, \cdots$ is a finite natural number and come back to Assumption A. The states $\tilde{\omega}_{k}^{\pm,T}$ are
(pure) states on the algebra ${\cal A}(\Lambda_{k})$ or ${\cal A}_{k}$, identified as algebras of bounded operators ${\cal B}({\cal H}_{k})$
(on ${\cal H}(\Lambda_{k})$ or ${\cal H}_{k}$) corresponding to the vectors $\Psi^{k,\pm,T}$. For simplicity of notation, let
$x_{k}^{T} \equiv \Psi^{k,+,T}$, $y_{k}^{T} \equiv \Psi^{k,-,T}$, ${\cal A}_{k}$ stands for ${\cal A}(\Lambda_{k})$ or ${\cal A}_{k}$,
${\cal H}_{k}$ for both ${\cal H}(\Lambda_{k})$ or ${\cal H}_{k}$, $\tilde{\omega}_{k}^{+,T} = \omega_{x_{k}^{T}}$, 
$\tilde{\omega}_{k}^{-,T} = \omega_{y_{k}^{T}}$. As usual,
\begin{equation}
\label{(2.28)}
||\omega_{x_{k}^{T}}-\omega_{y_{k}^{T}}|| = \sup_{A \in {\cal A}_{k}, ||A||\le 1} |\omega_{x_{k}^{T}}(A)-\omega_{y_{k}^{T}}(A)|
\end{equation}
but
\begin{equation}
\label{(2.29)}
\omega_{x_{k}^{T}}(A)-\omega_{y_{k}^{T}}(A) = (x_{k}^{T}, A x_{k}^{T}) - (y_{k}^{T},A y_{k}^{T})= tr_{{\cal H}_{k}}(T_{k} A)
\end{equation}
where
\begin{equation}
\label{(2.30)}
T_{k} \equiv x_{k}^{T} \otimes \overline{x_{k}^{T}} - y_{k}^{T} \otimes \overline{y_{k}^{T}} 
\end{equation}
with the definition
\begin{equation}
\label{(2.31)}
(x_{k} \otimes \overline{x_{k}})f \equiv (x_{k},f) x_{k} \mbox{ for } f \in {\cal H}_{k}
\end{equation}
Clearly, $T_{k}$ is an operator of rank 2, and therefore in the trace class, denoted $\tau c$ as in \cite{Sch}, and we have
(\cite{Sch}, Theorem 2, p.47)

\begin{lemma}
\label{lem:3.1}
The expression \eqref{(2.29)} represents a bounded linear functional on $\tau c$ of norm $||A||$. Moreover, $(\tau c)^{*}$
and ${\cal B}({\cal H})$ are equivalent, in the sense of Banach identical.
\end{lemma}

By the second assertion of Lemma ~\ref{lem:3.1},
\begin{equation}
\label{(2.32)}
||\omega_{x_{k}^{T}}-\omega_{y_{k}^{T}}|| = tr_{{\cal H}_{k}} (|T_{k}|)
\end{equation}
where $|T_{k}| \equiv (T_{k}^{\dag}T_{k})^{1/2}$. the eigenvalues of $|T_{k}|$ equal the absolute values of those of $T_{k}$; 
by \eqref{(2.30)}, \eqref{(2.31)} the latter may be obtained directly from the trace and determinant of the anti-Hermitian
matrix

\[ \left( \begin{array}{ccc}
1 & (x_{k},y_{k})\\
-(y_{k},x_{k}) & -1 \end{array} \right)\]

and equal
\begin{equation}
\label{(2.33a)}
\lambda_{1,k} = \sqrt(1-|(x_{k}^{T}, y_{k}^{T})|^{2})
\end{equation}
\begin{equation}
\label{(2.33b)}
\lambda_{2,k} = -\sqrt(1-|(x_{k}^{T}, y_{k}^{T})|^{2})
\end{equation}
Putting together \eqref{(2.32)} and \eqref{(2.33a)}, \eqref{(2.33b)}, we obtain the

\begin{corollary}
\label{cor:3.1}

\begin{equation}
\label{(2.34)}
|(x_{k}^{T}, y_{k}^{T})|^{2} = 1 - \frac{1}{4} ||\omega_{x_{k}^{T}}-\omega_{y_{k}^{T}}||^{2}
\end{equation}

\end{corollary}

Equation \eqref{(2.34)} suggests the natural definition, adapted from (\cite{RRoe}, Def. 4.7) to the present context:

\begin{definition}
\label{Definition 3.1}

Let, in the weak* topology,
\begin{equation}
\label{(2.35a)}
\omega_{x_{k}^{T}} \to \omega_{1}^{T}
\end{equation}
and
\begin{equation}
\label{(2.35b)}
\omega_{y_{k}^{T}} \to \omega_{2}^{T}
\end{equation}

The \emph{transition probability} between the states $\omega_{1}^{T}$ and $\omega_{2}^{T}$ on the C*-algebra ${\cal A}$, denoted
$\omega_{1}^{T}.\omega_{2}^{T}$, is defined as
\begin{equation}
\label{(2.35c)}
\omega_{1}^{T}.\omega_{2}^{T} \equiv \lim_{k \to \infty} (1 - \frac{1}{4}||\omega_{x_{k}^{T}} - \omega_{y_{k}^{T}}||^{2})
\end{equation}
whenever the limit on the r.h.s. of \eqref{(2.35c)} exists.

\end{definition}

We have the

\begin{theorem}
\label{th:3.1}

If the states $\omega_{1}^{T}$ and $\omega_{2}^{T}$ in Definition ~\ref{Definition 3.1} are disjoint (Definition ~\ref{Definition 2.1}), 
the transition probability between them is zero.

\begin{proof}

Considering the C* algebra ${\cal A}$ as a Banach space relatively to the weak topology on the dual space of states (the weak* topolgy),
the norm is lower semi-continuous (see, e.g., \cite{Cho}, Ex. 60, p.287), and thus \eqref{(2.35a)} and \eqref{(2.35b)} imply that
\begin{equation}
\label{(2.36)}
\liminf_{k \to \infty} ||\omega_{x_{k}^{T}} -\omega_{y_{k}^{T}}|| \ge ||\omega_{1}^{T} - \omega_{2}^{T}||
\end{equation}
Since $\omega_{1}^{T}$ and $\omega_{2}^{T}$ are disjoint, by the theorem of Glimm and Kadison \cite{GK}
\begin{equation}
\label{(2.37)}
||\omega_{1}^{T} - \omega_{2}^{T}|| = 2
\end{equation}
We thus have
\begin{eqnarray*}
0 \le \liminf_{k \to \infty} (1 - \frac{1}{4}||\omega_{x_{k}^{T}} -\omega_{y_{k}^{T}}||^{2}) \le \\
\limsup_{k \to \infty} (1 - \frac{1}{4}||\omega_{x_{k}^{T}} -\omega_{y_{k}^{T}}||^{2}) \le 0
\end{eqnarray*}

The first inequality above follows from the uniform bound $||\omega_{x_{k}^{T}} - \omega_{y_{k}^{T}}|| \le 2$ and the third inequality
above is a consequence of \eqref{(2.36)}. The assertion follows.

\end{proof}

\end{theorem}   

\begin{remark}
\label{Remark 3.1}

In (\cite{Hepp}, Lemma3, p.24) it was wrongly asserted that the norm is weakly continuous; the rest of his Lemma 3 contains, however,
an important idea, which we now use. Let $A \in {\cal A}$. If $\omega_{y_{k}^{T}}(A^{\dag}A) = 0$,
$$
|(\Psi^{k,-,T}, \pi_{k}(A) \Psi^{k,+,T})|^{2} \le \omega_{y_{k}^{T}}(A^{\dag}A) \to 0 \mbox{ as } k \to \infty
$$
Otherwise, $\omega_{y_{k}^{T}} (A^{\dag}A) \ne 0$ for $k$ sufficiently large, and we may define the state
\begin{eqnarray*}
\omega_{y_{k}^{T}}^{A} \equiv \\
\frac {(\Psi^{k,-,T}, \pi_{k}(A)^{\dag} \cdot \pi_{k}(A) \Psi^{k,-,T})}{(\Psi^{k,-,T}, \pi_{k}(A)^{\dag} \pi_{k}(A) \Psi^{k,-,T})}
\end{eqnarray*}
By \eqref{(2.35b)},
$$
\omega_{y_{k}^{T}}^{A} \to \omega_{2}^{T,A}
$$
in the weak* topology, where, by \eqref{(2.12)}, $\omega_{2}^{T,A} \in [\omega_{2}^{T}]$, the \emph{folium} of $\omega_{2}^{T}$  (defined
by (23).

\end{remark}

From the above, and the remarks following Lemma ~\ref{lem:1}, $\omega_{1}^{T}$ and $\omega_{2}^{T,A}$ are likewise disjoint, 
by the assumption of Theorem ~\ref{th:3.1}, implying the following

\begin{corollary}
\label{cor:3.1}
Under the same assumptions of Theorem ~\ref{th:3.1}, the transition probability between $\omega_{1}^{T}$ and $\omega_{2}^{T,A}$ is
zero for any $A \in {\cal A}$. In particular, by \eqref{(2.34)},
\begin{equation}
\label{(2.37)}
\lim_{k \to \infty} (\Psi^{k,+,T}, \pi_{k}(A) \Psi^{k,-,T}) = 0 
\end{equation}

\end{corollary}

\begin{remark}
\label{Remark 3.2}

Corollary ~\ref{cor:3.1} makes precise the replacement of \eqref{(1.10.1)} and \eqref{(1.10.2)} by \eqref{(1.11)} ``in the limit $N \to \infty$'', which
corresponds to the fact that the transition probability between the states $\omega_{1}^{T}$ and $\omega_{2}^{T,A}$ of the infinite
system is zero, for any $A \in {\cal A}$, according to Definition ~\ref{Definition 3.1}.

\end{remark}

In general, the disjointness of the two states in the assumption of Theorem ~\ref{th:3.1} is not easy to prove. In the next section,
we describe a model of Stern-Gerlach type in which two different proofs of this property may be given, as long as the time-of-measurement
parameter $T$ satisfies \eqref{(1.5)}. The second proof will relate disjointness to the values taken by the limiting states on classical
or macroscopic observables of type \eqref{(1.8)}, i.e., the ``pointer positions'' in measurement theory. 

\section{Application to a model of the Stern-Gerlach experiment}

\subsection{The model}

We describe in this section a model of the Stern-Gerlach experiment \cite{StGer}. A jet of silver atoms cross a strongly inhomogeneous 
magnetic field directed along the z-axis. We use the setting of Gondran and Gondran \cite{GG}, in which silver atoms of spin one-half
contained in an oven are heated to high temperature and escape through a narrow opening. A collimating fence $F$ selects those atoms whose
velocities are parallel to the y axis: it is assumed to be much larger along Ox, in such a way that both variables x and y may be treated 
classically. The atomic jet arrives then at an electromagnet at the initial time $t=0$, each atom being then described by the wave function
\begin{equation}
\label{(4.1)}
\Psi_{T}(z) = \Psi_{C}(z) (\alpha |+) + \beta |-))
\end{equation}
with $|\alpha|^{2}+|\beta|^{2} = 1$, $\sigma_{z}|\pm) = \pm |\pm)$, and the configurational part $\Psi_{C}$ is given by
\begin{equation}
\label{(4.2)}
\Psi_{C}(z) \equiv (2 \pi \sigma_{0}^{2})^{-1/2} \exp(\frac{-z^{2}}{4\sigma_{0}^{2}})
\end{equation}
After leaving the magnetic field, there is free motion until the particle reaches a screen placed beyond the magnet, at a certain time
$T$, when the measurement is performed.

We shall assume that each spin eigenstate is attached not only to one atom, but to all those atoms in a tiny neighborhood of a point
in space (e.g., of diameter of a micron), but still containing a macroscopic number $N$ of atoms. The Hamiltonian \eqref{(2.18)} is
thus assumed to be
\begin{equation}
\label{(4.3)}
H_{N} = H_{S} \otimes \mathbf{1} + \mathbf{1} \otimes H_{A_{k}} + V_{k}
\end{equation}
with
\begin{equation}
\label{(4.4)}
H_{S} = \mu \sigma_{z} B
\end{equation}
\begin{equation}
\label{(4.5)}
H_{A_{k}} = \frac{(P_{z}^{(k)})^{2}}{2 M_{k}}
\end{equation}
\begin{equation}
\label{(4.6)}
V_{k} = \lambda P_{z}^{(k)} \sigma_{z}
\end{equation}
with $M_{k} = (2k+1)m$, $m$ being the mass of a single atom, and
\begin{equation}
\label{(4.7)}
P_{z}^{(k)} = p_{z}^{-k}+ \cdots + p_{z}^{k}
\end{equation}
Note that we have replaced $N$ by $2k+1$, the integer variable runs from $-k$ to $k$, in order to have a model on $\mathbf{Z}$. 
The corresponding effective quantum spin model of the next subsection will be thereby a translation invariant model on the lattice 
$\mathbf{Z}$. The operator $p_{z}^{k}$ corresponding to each atom is the usual self-adjoint z-component of the momentum operator
acting on the Hilbert space $L^{2}(\mathbf{R})$, and the algebra, the one-dimensional Weyl algebra corresponding to the sole
variable z. Since, by \eqref{(4.6)}, each spin couples only to the z-component of the center of mass momentum, the corresponding
macroscopic operator will be the z-component of the center of mass coordinate $\frac{z_{-k} + \cdots + z_{k}}{2k+1}$ or, as we
shall see, the limit, for $\rho$ real
\begin{equation}
\label{(4.8)}
\lim_{k \to \infty} \exp(i\rho \frac{z_{-k} + \cdots + z_{k}}{2k+1})
\end{equation}
which will be seen to exist in the appropriate representation. The model \eqref{(4.3)}-\eqref{(4.7)} is an adaptation (to a version
of infinite number of degrees of freedom) of the model in the book by Gottfried and Yan (\cite{KG1}, pp. 559 et seq.). Equation
\eqref{(4.4)} represents the interaction with the constant part of the magnetic field, \eqref{(4.5)} the kinetic energy and
\eqref{(4.6)} the interaction with the field gradient, assumed to be along the z-direction

Since $H_{S}$ and $H_{A_{k}}$ commute with $V_{k}$, there is no problem in taking them into account, but that will only be an unnecessary
burden, which only changes some constants in the forthcoming account; consequently, we ignore them both (alternatively, take $m \to \infty$
and $B=0$). Thus our Hamiltonian will be
\begin{equation}
\label{(4.9)}
H_{k} = V_{k} = \lambda P_{z}^{(k)} \otimes \sigma_{z}
\end{equation}  

Before going on, we should like to explain the relation of the present model to the standard SG model-experiment in greater detail.

The Hamiltonian of the flying atoms should be
$$
H_{S} = \frac{p^{2}}{2m} + \mu \sigma_{z} B_{z}(z) = \frac{p^{2}}{2m} +  \mu \sigma_{z}( B_{z}(0) + z \frac{\partial B_{z}}{\partial z})
$$
However, from $\nabla \cdot \vec{B} = 0$, it follows that other components of the magnetic moment interact with the field, ``a fact that
is often ignored in text-book descriptions'', as remarked by Gottfried and Yan (\cite{KG1}, p. 558, bottom). They also remark 
that, as this issue is irrelevant to their purpose, they avoid it completely by constructing a soluble model that produces the same
results as a good SG experiment. This is the model we use in this chapter, but with the following additions and modifications.

First, we do not need to ignore the condition $\nabla \cdot \vec{B} = 0$, and assume that the particle first enters an electromagnetic
field $\vec{B}$ directed along the $z$ axis given by
$$
B_{x} = B^{'}_{0}x \mbox{ with } B_{y}=0 \mbox{ and } B_{z} = B_{0} - B^{'}_{0} z
$$
We employ the approximation 
$$
B^{'}_{0} = |\frac{\partial B_{z}}{\partial z}| = \mbox{ constant }
$$
Such a vector $\vec{B}$ does satisfy the Maxwell equation $\nabla \cdot \vec{B} = 0$.

Reference (\cite{GG} is one of the very few in which the \emph{spatial extension} of the spinor is
taken into account. This is, however, precisely the crucial element allowing to take into account the initial position $(x_{0},z_{0})$
of the particle and render the evolution of the quantum system deterministic: if it is eliminated, one loses the possibility of
individualizing the particle and, finally, to perform the measurement of the coordinate $z$ of the spots on the screen. Assuming that
the initial state of the silver atom is a bound state, a corresponding natural simplified Ansatz for it is a Gaussian
$$
\Psi_{0}(x,z) = (2\pi \sigma_{0}^{2})^{-1/2} \exp(-\frac{z^{2}+x^{2}}{4\sigma_{0}^{2}}) S
$$
where

\[ S=  \left( \begin{array}{c}                                                            
           \cos(\frac{\theta_{0}}{2})\exp(i\phi_{0}/2)\\                                               
            i\sin(\frac{\theta_{0}}{2})\exp(-i\phi_{0}/2) \end{array} \right) \] 

The solution of the time-dependent Schr\"{o}dinger equation for the spinor $\Psi$
$$
i\hbar \frac{\partial \Psi}{\partial t} = -\frac{\hbar^{2}}{2m}\nabla^{2} \Psi + \mu_{B} \vec{B}\cdot \vec{\sigma} \Psi
$$
with the above initial condition, the magnetic field $\vec{B}$ as given above, is the same as the solution obtained with the
Hamiltonian (65), see (3) of \cite{GG} and Appendix A of \cite{GG}. This is not unexpected because the multiplication operator $z$
acting on a Gaussian is equivalent to a derivation. This shows that our model is indeed the SG model ``in disguise''.

The silver atoms form a jet with a certain, nonzero finite density $\rho$. Their number $N$, in a macroscopic volume $V$, may be 
supposed to be well described by the thermodynamic limit $N \to \infty$, $V \to \infty$, $\frac{N}{V} = \rho$.
Since the $z$ coordinates of the two spots on the screen, in the SG experiment, are macroscopic numbers, it is reasonable to assume,
correspondingly, that they are obtained as mean values of (microscopic) averages of $z$ coordinates $z_{1}, \cdots, z_{N}$, i.e.,
$\lim_{N \to \infty} \frac{z_{1} + \cdots +z_{N}}{N}$. The external magnetic field gradient (supposed to be a constant equal to $\lambda$)
is also macroscopic and, accordingly, it seems reasonable to assume that
$$
\lambda (\sigma^{1}_{z} \otimes (z_{1} + \cdots + \sigma^{N}_{z}\otimes z_{N}) \approx \lambda \sigma_{z} \otimes (z_{1} + \cdots +z_{N})
$$
in a tiny (e.g. of the diameter of a micron) but still macroscopic vicinity of a point in configuration space. As explained, we
may replace $z_{1} + \cdots + z_{N}$ by $p_{1} + \cdots + p_{N}$, where $p_{i}$ denote momentum operators of the i-th particle.

Thus, the measurement, here ``performed'' by the coordinate wave-function, is ``arbitrarily close'' to one in a finite volume 
$V_{0}$, and the elements of the quasi-local algebra ${\cal A}$, which are arbitrarily close (in norm) to an element localized in 
a finite volume $V_{0}$, will not be able to distinguish between two disjoint states, because they are ``macroscopically different'',
i.e., differ from one another by an infinite number of operations - e.g., by flipping an infinite number of spins in states of 
different mean magnetizations, or changing the coordinates of the particles in jets of different values of the mean (C.M.) coordinate.

The fact that the coupling is assumed to occur only with the center of mass momentum explains why only the free motion is relevant
in the final formulas (see Remark 4.1), and justifies restriction to product states, because the eventual (e.g. van der Waals) interactions
between the silver atoms is entirely negligible.

We now proceed with the treatment of the model (65).

In correspondence to \eqref{(4.2)}, the initial ($t=0$) configurational state is
\begin{equation}
\label{(4.10)}
\Psi_{C,k,0}(z_{-k}, \cdots, z_{k}) = (2\pi \sigma_{0}^{2})^{-1/2} \exp(\frac{-z_{-k}^{2}+ \cdots -z_{k}^{2}}{4\sigma_{0}^{2}})
\end{equation}
and the full $t=0$ wave-vector associated to \eqref{(4.1)} becomes
\begin{equation}
\label{(4.11)}
\Psi_{T,k,0} = (\alpha |+) + \beta|-)) \otimes \Psi_{C,k,0}
\end{equation}
in the Hilbert space ${\cal H}= \mathbf{C}^{2} \otimes \otimes_{i=-k}^{k}L_{i}^{2}(\mathbf{R})$, where $L_{i}^{2}(\mathbf{R}$ denotes 
the $i-th$ copy of $L^{2}(\mathbf{R})$ associated to the $k-th$ particle. Equation \eqref{(4.9)} then yields
\begin{equation}
\label{(4.13)}
\exp(-itH_{k}) \Psi_{T,k,0} = \alpha |+) \otimes \Psi^{k,-,t} + \beta |-) \otimes \Psi^{k,+,t}
\end{equation}
with
\begin{equation}
\label{(4.14a)}
\Psi^{k,+,t}(z_{-k}, \cdots, z_{k})= \Psi_{C,k,0}(z_{-k}-\lambda t, \cdots, z_{k} -\lambda t)
\end{equation}
together with 
\begin{equation}
\label{(4.14b)}
\Psi^{k,-,t}(z_{-k}, \cdots, z_{k})= \Psi_{C,k,0}(z_{-k}+\lambda t, \cdots, z_{k} +\lambda t)
\end{equation} 
In correspondence with \eqref{(4.13)}, the states $\omega_{x_{k}^{T}}$, $\omega_{y_{k}^{T}}$ defined before \eqref{(2.28)} become
\begin{equation}
\label{(4.15a)}
\omega_{x_{k}^{T}}(A) = (\Psi^{k,+,t}, \pi_{k}(A) \Psi^{k,+,t})
\end{equation} 
and
\begin{equation}
\label{(4.15b)}
\omega_{y_{k}^{T}}(A) = (\Psi^{k,-,t}, \pi_{k}(A) \Psi^{k,-,t})
\end{equation} 
where $A \in {\cal A}$, the infinite product of Weyl algebras defined in Assumption A. For this model the $t_{D}$ in \eqref{(1.5)}
may be explicitly computed: after t=0, the density splits into a sum of two Gaussians, which become separated as long as the distance
between their centers is larger than the widths of the two Gaussians, viz. $3\sigma_{0}$: $t_{D} = \frac{3\sigma_{0}}{\lambda}$ where
$\lambda$ stands for the average velocity in the $z$ direction: see (6) and (9) of \cite{GG} and the forthcoming \eqref{(4.27)}.

\begin{proposition}
\label{prop:4.1}

Let $T$ satisfy \eqref{(1.5)}. Then, the weak* limits of the states \eqref{(4.15a)}, \eqref{(4.15b)}, denoted by $\omega_{1}^{T}$ 
and $\omega_{2}^{T}$ as in Definition ~\ref{Definition 3.1}, are disjoint.

\begin{proof}

We are in the setting of Proposition ~\ref{prop:2.1}, with $x_{i} = \Psi^{i,+,T}$, on the one hand, and $y_{i} = \Psi^{i,-,T}$ on the other.
We have, by \eqref{(4.14a)}, \eqref{(4.14b)},
\begin{eqnarray*}
(x_{i},y_{i}) = (y_{i},x_{i}) = \\
= (2\pi \sigma_{0}^{2})^{-1} \int_{-\infty}^{\infty} dz_{i} \exp(-\frac{(z_{i}-\lambda T)^{2}}{4\sigma_{0}^{2}}) \times \\
\times \exp(-\frac{(z_{i}+\lambda T)^{2}}{4\sigma_{0}^{2}}) = \\
= (2\pi \sigma_{0}^{2})^{-1} \int_{-\infty}^{\infty} dz_{i}\exp(-\frac{z_{i}^{2}}{2\sigma_{0}^{2}}) \exp(-\frac{\lambda^{2}T^{2}}{2\sigma_{0}^{2}})=\\
= \exp(-\frac{\lambda^{2} T^{2}}{2\sigma_{0}^{2}})
\end{eqnarray*}
By \eqref{(1.5)}
$$ 
\exp(-\frac{\lambda^{2} T^{2}}{2\sigma_{0}^{2}}) \ge \exp(-\frac{\lambda^{2} t_{D}^{2}}{2\sigma_{0}^{2}}) 
$$ 
and hence
\begin{equation}
\label{(4.17)}
|(x_{i},y_{i}) -1| = 1-\exp(-\frac{\lambda^{2} t_{D}^{2}}{2\sigma_{0}^{2}}) \ge \frac{1}{4}\frac{\lambda^{2} t_{D}^{2}}{2\sigma_{0}^{2}}
\end{equation}
By Definition ~\ref{Definition 2.4}, the representations of ${\cal A}$ canonically associated to the infinite product states
$\omega_{1}$ and $\omega_{2}$ are $\otimes_{i \in \mathbf{Z}}^{\otimes\xi_{i}} \pi_{i}$, with $\xi_{i} = x_{i} \mbox{ or } y_{i}$, and
the corresponding $C_{0}$ - families are not equivalent by \eqref{(4.17)} and Definition ~\ref{Definition 2.3}, hence they are
disjoint.

\end{proof}

\end{proposition}

The fact used above that ``not not-equivalent'' means disjointness as defined by definition ~\ref{Definition 2.1} may not be 
immediately clear but it, too, follows from Lemma \ref{lem:1}. First, we dispose of $A$ because of quasi-locality, and, due
to the product structure, we arrive as a necessary and sufficient condition for disjointness of
states, that the scalar product $\prod_{i \in I; |i| sufficiently large} (x_{i},y_{i}) = 0$, or, taking the logarithm
$$
|\log(\prod_{i \in I; |i| sufficiently large} (x_{i},y_{i}))| = \infty
$$
In rigorous terms, this is replaced by the condition
$$
\sum_{i \in I; |i| sufficiently large} |(x_{i},y_{i})-1| = \infty
$$
This replacement is due to the necessity of avoiding the problems related to zero factors in the infinite product, or to ``infinite phases'',
see \cite{Wehrl}. The above condition may be intuitively motivated by the fact that \emph{convergence} of the infinite product implies
that each term must tend to one: considering the logarithm of the product, each $\log(x_{i},y_{i})$ is close to 
$1-(x_{i},y_{i})$ and, thus, convergence means 
$$
\sum_{i \in I; |i| sufficiently large} |(x_{i},y_{i})-1| < \infty
$$
of which the previous formula is the negation.

We come now to a second proof of disjointness, which both illuminates its physical content and defines precisely the results and
parameter values associated to the measurement. The (z-component of) the center of mass of the atoms \eqref{(1.8)} is
\begin{equation}
\label{(4.18)}
z_{C.M.} = \lim_{k \to \infty} \frac{1}{2k+1} \sum_{i=-k}^{k} z_{k}
\end{equation}
The above limit may be seen to exist in each IDPS $\otimes_{i \in \mathbf{Z}}^{\otimes \xi_{i}}$, with $\xi_{i}=x_{i} \mbox{ or } y_{i}$,
assuming different values in each representation:

\begin{proposition}
\label{prop:4.2}
$z_{C.M.}$ exists in the sense that, for any $\rho \in \mathbf{R}$,
\begin{equation}
\label{(4.19a)}
\lim_{k \to \infty} \exp(i \rho \frac{\sum_{i=-k}^{k} z_{k}}{2k+1}) = \exp(i\rho \lambda T)
\end{equation}
in the IDPS $\otimes_{i \in \mathbf{Z}}^{\otimes x_{i}}$, and
\begin{equation}
\label{(4.19b)}
\lim_{k \to \infty} \exp(i \rho \frac{\sum_{i=-k}^{k} z_{k}}{2k+1}) = \exp(-i\rho \lambda T)
\end{equation}
in the IDPS $\otimes_{i \in \mathbf{Z}}^{\otimes y_{i}}$. As a consequence, the two IDPS are disjoint.

\begin{proof}

We have
\begin{eqnarray*}
(\Psi^{k,+,T},\exp(i \rho \frac{\sum_{i=-k}^{k} z_{k}}{2k+1})\Psi^{k,+,T})= \\
= (2\pi \sigma_{0}^{2})^{-\frac{2k+1}{2}}\int_{-\infty}^{\infty} dz_{-k} \cdots \int_{-\infty}^{\infty} dz_{k} \\
 \exp(-2\frac{(z_{-k}-\lambda T)^{2}}{4\sigma_{0}^{2}}) \cdots  \exp(-2\frac{(z_{k}-\lambda T)^{2}}{4\sigma_{0}^{2}})\\
 \exp(i \rho \frac{z_{-k}}{2k+1}) \cdots \exp(i \rho \frac{z_{k}}{2k+1}) = \\
= \exp(-\frac{\rho^{2}\sigma_{0}^{2}}{2(2k+1)}) \exp(i \rho \lambda T)
\end{eqnarray*}
from which
\begin{equation}
\label{(4.20a)}
\lim_{k \to \infty} (\Psi^{k,+,T},\exp(i \rho \frac{\sum_{i=-k}^{k} z_{k}}{2k+1})\Psi^{k,+,T}) = \exp(i \rho \lambda T)
\end{equation}
and, analogously,
\begin{equation}
\label{(4.20b)}
\lim_{k \to \infty} (\Psi^{k,-,T},\exp(i \rho \frac{\sum_{i=-k}^{k} z_{k}}{2k+1})\Psi^{k,-,T}) = \exp(-i \rho \lambda T)
\end{equation}

By Proposition ~\ref{prop:2.1} and the fact that the limits on the left hand sides of \eqref{(4.20a)}, \eqref{(4.20b)} 
are not altered by changing the variables $z_{i}$ with $i$ in a finite set we may replace $\Psi^{k,\pm,T}$ in equations
\eqref{(4.20a)}(resp. \eqref{(4.20b)}) by vectors in a total set in  $\otimes_{i \in \mathbf{Z}}^{\otimes x_{i}}$
(resp.$\otimes_{i \in \mathbf{Z}}^{\otimes y_{i}}$). This shows \eqref{(4.19a)} and \eqref{(4.19b)}. Disjointness of
the IDPS is a consequence of an argument identical to the one used in connection with \eqref{(2.14)}.

\end{proof}

\end{proposition}

\subsection{An effective quantum spin model}

An effective quantum spin model for the previously studied Stern-Gerlach model is obtained by replacing 
$\otimes_{-k}^{k} L_{i}^{2}(\mathbf{R})$ by  

$$
{\cal H}_{k} = \otimes_{i=-k}^{k} \mathbf{C}_{k}^{2}
$$
Given a fixed $T$ satisfying \eqref{(1.5)}, perform in the states $\omega_{x_{k}^{T}}$, $\omega_{y_{k}^{T}}$ in \eqref{(4.15a)}
and \eqref{(4.15b)} the substitution
\begin{equation}
\label{(4.22)}
\Psi^{k,\pm,T} \to \otimes_{i=-k}^{k} |\pm)_{k}
\end{equation}
where $|\pm)_{k}$ are, as before, the spin eigenstates of $\sigma^{z}_{k}:\sigma^{z}_{k}|\pm)_{k} = \pm |\pm)_{k}$, together with
the substitution
\begin{equation}
\label{(4.23a)}
\lim_{k \to \infty} \exp(i \rho \frac{\sum_{i=-k}^{k} z_{k}}{2k+1}) \to \lim_{k \to \infty} \exp(2i\rho T \frac{\sum_{i=-k}^k \sigma^{z}_{i}}{2k+1})
\end{equation}

Then: the weak* limit of the sequence of states
\begin{equation}
\label{(4.23b)}
\omega_{k} \equiv |\alpha|^{2} \omega_{k}^{+} + |\beta|^{2} \omega_{k}^{-}
\end{equation}
with $|\alpha|^{2} + |\beta|^{2}= 1$, on the quasi-local algebra ${\cal A}$ associated to the spin algebra on $\mathbf{Z}$
and $\omega_{k}^{\pm}$ denoting the product states which define the familiar disjoint representations $\pi_{\vec{m}}$, $\pi_{\vec{n}}$
(with $\vec{m} = \pm (0,0,1)$) described in section 2, after (23), is an \emph{effective} quantum spin model for the SG model described 
in the previous subsection, in the sense that it reproduces the ``quantities to be measured'' \eqref{(4.19a)}, \eqref{(4.19b)}, as
long as the substitution \eqref{(4.23a)} is performed.

The present model serves as illustration of the remarks on irreversibility in the next section 5.

\begin{remark}
\label{Remark 4.1}
It is of course critical that $t_{D} \ne 0$ in \eqref{(1.5)}; the case of ``instantaneous measurement'' is excluded by the
Basdevant-Dalibard assumption a). The same requirement is independently imposed by the theory of irreversibility, see the next section.

As a further concrete illustration of this requirement in the present model, note that by the equations preceding \eqref{(4.17)},
\begin{equation}
\label{(4.24)}
(\Psi^{k,+,T}, \Psi^{k,-,T}) = \exp(-\frac{\lambda^{2} T^{2} k}{2\sigma_{0}^{2}})
\end{equation}
so that, if
\begin{equation}
\label{(4.25)}
T = T(k) = O(\frac{1}{\sqrt(k)})
\end{equation}
the cross terms in \eqref{(2.19b)} do not tend to zero. The fact that the possibility $T(k) \to 0$ as $k \to \infty$ is to be
\emph{excluded}, contrarily to the remarks in \cite{Dop}, has a simple explanation, to be given next.

Finally, it should be remarked that equation \eqref{(4.24)} shows explicitly that the condition $k \to \infty$ is not always
\emph{necessary} to achieve a very high degree of decoherence. Indeed, let $T=1 sec$ and $\frac{\sigma_{0}}{\lambda} = 10^{-4} sec$
(the latter reasonable experimental values, see (9) in \cite{GG}), and $k=1$ (i.e., just one
particle), we obtain for the r.h.s. of equation \eqref{(4.24)} the value $\exp(-10^{8})$, a forbiddingly small value! 

\end{remark}

\begin{remark}
\label{Remark 4.2}

If we differentiate equations \eqref{(4.19a)} and \eqref{(4.19b)} with respect to $\rho$, setting $\rho=0$ afterwards, we obtain,
denoting $<z_{C.M}>_{T}$ the expectation of the C.M. variable (74) in the product state at time $T$:

\begin{equation}
\label{(4.27)}
<z_{C.M.}>_{T} = \pm \lambda T = 2 s_{z} \lambda T
\end{equation}
where
\begin{equation}
\label{(4.28)}
s_{z} = \pm \frac{1}{2}
\end{equation}
are the two values of the z component of the spin operator, which comprise, in this experiment, the ``measured values''. Equation
\eqref{(4.27)} is essentially equation (65) of p. 559 of \cite{KG1} (with $P_{z}=0$, which we assumed) - not surprisingly the 
solution of the classical equation of motion, because the Gaussians are coherent states.

By \eqref{(4.27)}, $T$ is proportional to the value of a ``macroscopic observable'' $<z_{C.M.}>_{T}$, independent of $k$. This explains
why a behavior such as \eqref{(4.25)} is excluded, or, more generally, that the possibility $T(k) \to zero$ as $k \to \infty$
mentioned in \cite{Dop} is excluded.

The two values \eqref{(4.28)} are obtained from \eqref{(4.27)} through the measured values of $<z_{C.M.}>_{T}$ and $T$ (with a known constant
$\lambda$) and remain constant when the ``observer'' $(<z_{C.M.}>_{T},T)$ changes; the ``intrinsic property postulate'' of Bell and Gottfried
is therefore verified in the present model.

Finally, the mathematical limit $T \to \infty$ is unphysical in this model, since it corresponds to place the screen at infinite
distance from the electromagnet.

\end{remark}

\section{Irreversibility, the time-arrow and the conservation of entropy under measurements}

In his conclusion, Hepp \cite{Hepp} remarks: ``The solution of the problem of measurement is closely connected with the yet unknown
correct description of irreversibility in quantum mechanics''.

One such description of closed systems, without changing the Schr\"{o}dinger equation and the Copenhagen interpretation was
proposed in \cite{Wre}, see also \cite{Wre1} for a comprehensive review, which includes the stability of the second law in
the form proposed in \cite{Wre} under interactions with the environment.

For a finite quantum spin system the Gibbs-von Neumann entropy is ($k_{B}=1$)
\begin{equation}
\label{(5.1)}
S_{\Lambda} = -Tr (\rho_{\Lambda} \log \rho_{\Lambda})
\end{equation}
As remarked in section 2, we may view $\rho_{\Lambda}$ as a state $\omega_{\Lambda}$ on ${\cal A}(\Lambda)$ which generalizes to systems with
infinite number of degrees of freedom $\omega(A)= \lim_{\Lambda \nearrow \infty} \omega_{\Lambda}(A)$, at first for $A \in {\cal A}_{L}$
and then to ${\cal A}$. For a large system the \emph{mean entropy} is the natural quantity from the physical standpoint:
\begin{equation}
\label{(5.2)}
s(\omega) \equiv \lim_{\Lambda \nearrow \infty} (\frac{S_{\Lambda}}{|\Lambda|})(\omega)
\end{equation}
The mean entropy has the property \cite{LanRo}:
\begin{equation}
\label{(5.3)}
0 \le s(\omega) \le \log D \mbox{ where } D=2S+1 
\end{equation}
where $S$ denotes the value of the spin, in the present paper and in the effective model of section 4.2, $S=\frac{1}{2}$.

In his paper ``Against measurement'', John Bell, in a statement which is qualitatively similar to Hepp's, insisted on the 
necessity of physical precision regarding such words as reversible, irreversible, information (whose information? information about what?).

The theory developed in \cite{Wre}, \cite{Wre1} starts defining an adiabatic transformation, in which there is a first step, 
a finite preparation time $t_{p}$, during which external forces act, at the end of which the Hamiltonian associated to the initial 
equilibrium state is restored, and remains so ``forever'' during the second step. In measurement theory, Lamb \cite{Lamb} also
emphasizes the dual role of preparation and measurement. If the time of measurement $T$ is such that $T > t_{p}$, and the wave-vector
describing the system  is not identically zero in the whole interval $[0,T]$, the dynamics of the system in the time interval
$[-t_{r},T]$, of preparation followed by measurement, is *not* time-reversal invariant, leading to a \emph{time arrow}. If $t_{p}=T=0$,
i.e., both preparation and measurement are instantaneous, no guarantee of the existence of a time-arrow can be given. 

According to our theory, given a time arrow, the process $\omega_{1}(0) \to \omega_{2}(\infty)$ is defined to be reversible (irreversible) 
iff the inverse process $\omega_{2}(0) \to \omega_{1}(\infty)$ is possible (impossible). The first alternative takes place iff 
$s(\omega_{1}) = s(\omega_{2})$, the second one iff $s(\omega_{1}) < s(\omega_{2})$. Infinite time $t=\infty$ means, physically, that $T$ is
much larger than a quantity $t_{D}$, the decoherence time, as explained in section 1.

Of course, irreversibility is incompatible with time-reversal invariance, because the mean entropy cannot both strictly increase 
and strictly decrease with time. This is a precise wording in our framework of the Schr\"{o}dinger paradox \cite{Schr}, cited in Lebowitz's 
inspiring review of the issue of time-assymetry \cite{Leb}.

We know that the space of states is convex and the entropy of a finite system satisfies the inequality ($0 \le \alpha  \le 1$)
\begin{equation}
\label{(5.4)}
S_{\Lambda}(\alpha \rho_{\Lambda}^{1}+(1-\alpha) \rho_{\Lambda}^{2}) > \alpha S_{\Lambda}(\rho_{\Lambda}^{1}) + (1-\alpha) S_{\Lambda}(\rho_{\Lambda}^{2})
\end{equation}
i.e, $S_{\Lambda}$ is \emph{strictly concave}: entropy is gained by mixing, but the gain is *not* extensive and disappears upon division
by $|\Lambda|$ and taking the infinite volume limit (inequalities of Lanford and Robinson \cite{LanRo}), so that the mean entropy becomes \emph{affine}:
\begin{equation}
\label{(5.5)}
s(\alpha \omega_{1} + (1-\alpha) \omega_{2}) = \alpha s(\omega_{1}) + (1-\alpha) s(\omega_{2})
\end{equation}

The state \eqref{(2.19b)} , \eqref{(2.19c)} tends, in the weak* topology, to a state $\omega_{T}$ (now on the algebra of system and apparatus
\eqref{(2.20)}, whose Gibbs-von Neumann entropy
is identical to that of the initial state $\omega_{0}$, and equals zero since the state is pure. The associated mean entropy 
therefore also satisfies
\begin{equation}
\label{(5.6)}
s(\omega_{T}) = s(\omega_{0})=0
\end{equation}

By Theorem ~\ref{th:3.1}, the state $\omega_{T}$  is equivalent, ``for all observables found in Nature'' to the ``collapsed state'' 
$\omega_{C}$ given by
\begin{equation}
\label{(5.7)}
\omega_{C} \equiv |\alpha|^{2}\omega_{1}^{T} + |\beta|^{2} \omega_{2}^{T}
\end{equation}
where, by \eqref{(2.19b)}, \eqref{(2.19c)}, \eqref{(2.21)}, \eqref{(2.22)}
\begin{equation}
\label{(5.8a)}
\omega_{1}^{T} = \omega_{T}(P_{+} \cdot)(|\alpha|^{2})^{-1}
\end{equation}
and
\begin{equation}
\label{(5.8b)}
\omega_{2}^{T} = \omega_{T}(P_{-} \cdot)(1-|\alpha|^{2})^{-1}
\end{equation}
with the notation $P_{+}=|+)(+|$, $P_{-}=|-)(-|$, the familiar projectors on the two eigenstates of $\sigma_{z}$.

\begin{theorem}
\label{th:5.1}
On the average the mean entropy is conserved by measurements, and remains equal to zero.

\begin{proof}

On the average, the mean entropy equals
\begin{eqnarray*}
|\alpha|^{2} s(\omega_{1}^{T}) + (1-|\alpha|^{2}) s(\omega_{2}^{T}) = \\
= s(|\alpha|^{2}\omega_{1}^{T} + (1-|\alpha|^{2})\omega_{2}^{T}) = \\
=s(\omega_{T}(P_{+} \cdot) + \omega_{T}(P_{-} \cdot)) =s(\omega_{T})=\\
=s(\omega_{0})= 0 
\end{eqnarray*}

The first equation above is due to the property of affinity \eqref{(5.5)}, the second one follows from \eqref{(5.8a)}, \eqref{(5.8b)},
the third one by the linearity of the states, and the fact that $P_{+}+P_{-} = \mathbf{1}$, the fourth from \eqref{(5.6)}.

\end{proof}
\end{theorem}

The above theorem relies on the property of affinity of the mean entropy, which has only been proved for quantum lattice systems \cite{LanRo}.
The sole example we are able to give is the effective quantum spin model of the Stern Gerlach experiment of section 4.1 which was given
in section 4.2. 

In contrast to the behavior found in Theorem ~\ref{th:5.1}, the Boltzmann and Gibbs-von Neumann entropy of a finite system 
is \emph{reduced} under collapse, by Lemma 3 of \cite{NarWre}. This may be understood as follows. 
Entropy $S_{\Lambda} = |\Lambda|\log D -I_{\Lambda}$, with $I_{\Lambda}$ denoting the (quantum) information.
For quantum spin systems $0 \le S_{\Lambda}/|\Lambda| \le \log D$, and therefore $0 \le I_{\Lambda}/|\Lambda| \le \log D$. It attains
its maximum value for pure states, which are characterized by $S_{\Lambda} = 0$. Under ``collapse'', each collapsed state is pure
and therefore information is gained: this explains that the (Boltzmann and von Neumann) entropies are reduced, on the average,
violating the second law (on the average). If one chooses to define irreversibility in terms of the growth of the 
quantum Boltzmann entropy, we arrive at the necessity, commented in the last paragraph of \cite{NarWre}, that interactions with 
the environment (as well as measurements) must be rare phenomena on the thermodynamic scale in order to account for the validity 
of the version of the second law which was proved in \cite{NarWre}. Our approach through the mean entropy seems therefore particularly 
natural in this context, and has the following physical interpretation. Equivalently to the previously discussed informational content (for
quantum spin systems), entropy is, in Boltzmann's sense, a measure of a macrostate's wealth of ``microstates'', and therefore grows by mixing, 
but it turns out that this growth is *not* extensive and disappears upon division by $|\Lambda|$ , i.e., taking the infinite volume limit 
(inequalities of Lanford and Robinson \cite{LanRo}), so that the affinity property \eqref{(5.5)} results and, with it, Theorem ~\ref{th:5.1}, 
confirming, \emph{in the sense of mean entropy}, Nicolas van Kampen's conjecture \cite{vK} that \emph{the entropy of the Universe is not 
affected by measurements}. It is not affected either by more general interactions with the environment \cite{NTh1}, resulting in
the \emph{stability of the second law} proved in \cite{Wre}, see \cite{Wre1}.

\section{Conclusion and open problems}

One central and dominating feature of the analysis over finite vs infinite dimensional spaces is that in the infinite dimensional 
case the solution may depend \emph{discontinuously} on the parameters of the problem. Indeed, infinite systems may exhibit 
\emph{singularities}, not present in finite macroscopic systems, well-known in the theory of \emph{phase transitions}: they
are parametrized by \emph{critical exponents}, which, moreover, display \emph{universal} properties, in excellent agreement with experiment!
The crucial example of ``discontinuity'', as $N \to \infty$,  in the context of measurement theory, is the basic structural
change of the states: a sequence of pure states may tend to a mixed state, by Theorem ~\ref{th:3.1}, as a consequence of the property
of disjointness, which has no analogue for finite system: in measurement theory, the mean entropy is conserved and equals zero by
Theorem ~\ref{th:5.1}. It may also happen that a sequence of pure states of infinite systems, parametrized by
the time variable, tends to a mixed state of strictly higher mean entropy (\cite{Wre}, \cite{Wre1}). 

In greater generality, the physical ``N large but finite'' differs qualitatively from ``N infinite'' because the latter exhibits 
\emph{universal} properties not found in finite systems. One example of these universal properties, crucial in our approach, is the 
affinity of the mean entropy, whose finite-volume counterpart strict concavity of $\frac{S_{\Lambda}}{|\Lambda|}$) is not universal because, 
not being uniform in $|\Lambda|$, it depends on the volume $|\Lambda|$ of the system. The fact that (only) ``N infinite'' is in good agreement 
with experiment is explained by the fact that, with $N \approx 10^{24}$, macroscopic systems are extremely close to infinite systems 
(the success of the thermodynamic limit!). This explains why we are able to complement, and sometimes improve on, Sewell's approach
to the measurement problem in (\cite{Se1}, \cite{Se2}).

The above-mentioned universality in the framework introduced here and in (\cite{Wre}, \cite{Wre1}) suggests that other physical
theories besides quantum spin systems might exhibit similar properties, e.g., relativistic quantum field theory (rqft), and, from 
there, hopefully, nonrelativistic quantum continuous systems by the non-relativistic limit of rqft. Since, however, rqft deals with 
fields and thus continuous quantum systems, the structure of the space of states is quite different from that of quantum spin systems, 
and, in particular, the states must be required to be locally normal \cite{DDR} or locally finite (\cite{Se}, p.26) - of which the
only existing proof in an interacting field theory is due to Glimm and Jaffe, for the vacuum state \cite{GliJa}. Moreover, Theorem
~\ref{th:3.1} and its corollary show that in measurement theory the relevant state is (equivalent to) a weak* limit of a sequence of 
convex linear combinations of product states, that is, a non-entangled state, according to the Bertlmann-Narnhofer-Thirring geometrical
picture of entanglement (\cite{NTh2}, \cite{BNTh}). The latter \cite{NTh2}, however, also suggests that in rqft ``almost every state is
entangled'' in a precise sense, and, indeed, Summers and Werner \cite{SW} and Landau \cite{La} show that the vacuum state in rqft
maximally violates Bell's inequality (see also Wightman's review \cite{Wight1} of their work). It is thus expected that entanglement
will play a role in a future theory of measurement in rqft.

The formulation of a theory of measurement in rqft is a difficult, very fundamental open problem: it is formulated as Problem 4 in Wightman's list
\cite{Wight1}: ``to examine the effects of relativistic invariance on measurement theory'', see also \cite{Lamb}. In particular, 
Doplicher suggests \cite{Dop} that the apparent ``nonlocalizability'' of the type observed in the EPR thought experiment, 
due to the superposition principle, would certainly disappear if truly \emph{local} measurements were performed - and spin or 
angular momentum measurements are not such. In fact, instead of \eqref{(4.6)}, we must have a true interaction between fields. 
Incidentally, for interacting fields, the singularity hypothesis of \cite{JaWre} implies that fields are not defined for sharp times, 
and ``instantaneous measurements'' are excluded.

Since the measurement problem in quantum mechanics is a very complex and controversial problem, the complexity being partly due to the 
variety of the existent physical situations, it cannot be hoped that this paper contains a ``final solution'' to the measurement problem. 
In particular, almost perfect decoherence may occur even if the limit $N \to \infty$ is not performed at all, when the time $T$ of observation
is sufficiently large: an explicit example of this situation is given in Remark 4.1. 

The above-mentioned example relates to the work of Machida and Namiki \cite{MN}, commented by Araki \cite{Ar}, who formulated the 
Machida-Namiki theory in terms of continuous superselection rules. The reduction of the wave-packet proceeds, then, as a consequence
of the (mathematical) limit $T \to \infty$ (in our notation: see equations (3.4), (3.5) in \cite{Ar}). Although, as we have argued,
this limit need not, in general, be of physical relevance (and, indeed, it is not in the SG model, see the last sentence in Remark 4.2),
we have just seen that almost perfect decoherence may occur, nevertheless, if $T$ is sufficiently large with respect to the decoherence
time $t_{D}$. These theories, therefore, do remain of considerable interest as a complement to ours.

Another example is provided by the question of
whether it is possible to devise any experiment (of the Bell-EPR type) which simultaneously measures precise values of incompatible observables 
(the SG experiment of section 4 being not of this type). This may indicate a different route to the previous discussion based on microcausality.
See, in this connection, the specific analysis in \cite{Gri4}, as well as the more general \cite{Gri1}; both are based on Griffiths' 
(probabilistic) theory of consistent histories (\cite{Gri2}, see also \cite{Gri3}).
The latter theory was used by Omn\`{e}s \cite{Omn}, who proposed to consider  measurements specified by special kinds of history
in which decoherence results in the classical behavior of the macroscopic variables of the apparatus, to a sufficient approximation. This
(not necessarily perfect) decoherence is yet another alternative, complementary approach to ours. Concerning the classical, macroscopic observables, 
it is also of special interest that they are shown in \cite{Req} to be special cases of a subalgebra of the class of microscopic quantum observables 
of a generic many-body system (see also \cite{Lud}). 

In spite of the above-mentioned limitations, we believe that the universal properties of perfect decoherence, 
as described in the first two paragraphs,
suggest that it is relevant, in the sense of an idealized limit, to a significant number of physical measurements, in which 
both the ``Heisenberg paradox'' and the ``irreversibility paradox'' have been eliminated, and, therefore, 
quantum mechanics, in the original Copenhagen interpretation, is totally free of internal inconsistencies. This occurs, however, as
Hepp \cite{Hepp} predicted, only if one takes into account the extension of quantum mechanics to systems with an infinite number 
of degrees of freedom, as formulated in \cite{RRoe} and \cite{Hepp}, and developed in section 3. 

\emph{Acknowledgements} We should like to thank Pedro L. Ribeiro for discussions, the late Derek W. Robinson for a correspondence
in which he stressed the importance of the property of affinity, and Professor R. B. Griffiths for an enlightening correspondence on
related matters. This paper owes very much to the remarks of both referees, which resulted in truly substantial
improvements regarding the previous version.

\qquad
\textbf{Statement concerning data availability}
\textbf{The author confirms that all the data supporting the findings of this study are available within the article.}
\qquad

\end{document}